\documentclass[english]{article}
\usepackage{helvet}

\usepackage[T1]{fontenc}
\usepackage[latin9]{inputenc}

\makeatletter

\providecommand{\tabularnewline}{\\}

\makeatother

\usepackage{babel}
\usepackage{listings}

\begin{document}
\title{Immutability and Design Patterns in Ruby}
\author{Seamus Brady, seamus@corvideon.ie}
\date{28/03/2013}
\maketitle
\begin{abstract}
Functional Programming has seen a resurgence in interest in the last
few years and is often mentioned in opposition to Object-Orientated
Programming. However, Object-Orientated Programming can actually absorb
some of the lessons of Functional Programming to improve performance
and reliability. This paper examines three ways in which Immutability,
a common Functional Programming technique, can be used to develop
immutable objects in Ruby.
\end{abstract}

\section{Introduction}

Functional Programming has left it's mathematical, academic environment
and become much more visible in recent years. Languages such as Erlang,
Haskell and Clojure are being used in many corporate environments.\cite{1} 

Immutable data is a standard feature of Functional Programming \cite{2}.
Immutability, when applied to objects, creates objects that cannot
be modified after they have been created \cite{6}. Immutable Objects
have certain advantages over standard Mutable Objects. This paper
considers how a programmer using Ruby, a popular Object-Orientated
programming language, can create such Immutable Objects. 

Section 2 outlines the Functional Programming approach and how some
of it's advantages may be appropriated by Object-Orientated programmers. 

There are recommended strategies for creating immutable objects in
strongly typed languages such as Java \cite{5}. In Section 3 three
of these strategies have been reworked into ``Proto Design Patterns''
to fit Ruby's dynamically typed nature.

\subsection{Resources and Methods}

This problem was approached by researching existing advice for creating
Immutable Objects in Java and redefining the approaches suggested
to work with Ruby.

\section{Background}

\subsection{The Problem of Mutable Objects}

In Object-Orientated programming, objects contain both program state
(data) and the means to operate on that state (methods). This simple
encapsulation allows an Object-Orientated language to model complex
systems very efficiently.

However, this approach has two difficulties:
\begin{quote}
Mutable objects are not thread-safe by default and it\textquoteright s
easy for clients to change their state outside our control. \cite{7}
\end{quote}
In an era when more and more systems have multiple processors, thread
safe code is essential for certain areas of a system that demand concurrent
programming \cite{8}. Also, as systems get ever more complex, subtle
bugs caused by mutable objects in complex data structures are very
difficult to fix \cite{4}.

\subsection{Monkey Patching in Ruby}

Ruby is a general purpose Object-Orientated programming language \cite{12}.
As it is a dynamically typed ``scripting'' language it is often
used in preference to Java. The challenge in designing Mutable Objects
in Ruby is that, unlike Java, classes can be changed (``monkey patched'')
as any point in the program \cite{13}.

This means that Ruby is endlessly flexible, but monkey patching can
create very subtle and difficult to find bugs \cite{14}. 

If an object is being used in part of the system that demands thread
safety, or where unwanted changes in state could cause problems, using
an Immutable Object could help avoid bugs.

\subsection{Functional Programming Approach}

Functional Programming is a distinct programming approach that treats
a computer program as a series of functions to be evaluated: 
\begin{quote}
Functional programming is so called because a program consists entirely
of functions. The main program itself is written as a function which
receives the program\textquoteright s input as its argument and delivers
the program\textquoteright s output as its result.\cite{1}
\end{quote}
One of the main reasons that Functional Programming is recommended
by developers is that changes in the program's state are avoided as
part of the Functional approach:
\begin{quote}
...functional programs contain no side-effects at all. A function
call can have no effect other than to compute its result. This eliminates
a major source of bugs, and also makes the order of execution irrelevant
- since no side-effect can change the value of an expression, it can
be evaluated at any time. \cite{1}
\end{quote}
To a commercial programmer who spends too much time debugging problems
that arise from inadvertent changes to program state, this seems like
a dream come true \cite{4}. However, Functional Programming is often
dismissed as an viable alternative for two reasons (amongst others):
\begin{itemize}
\item Functional Programming has a reputation for being difficult \cite{3}.
\item Most developers have spent time and money investing in alternative
approaches to programming and may be reluctant to change to something
new.
\end{itemize}

\subsection{Immutable Objects By Default}
\begin{quote}
Classes should be immutable unless there's a very good reason to make
them mutable....If a class cannot be made immutable, limit its mutability
as much as possible. \cite{9}
\end{quote}
However, Object-Orientated programmers do not have to take up Functional
Programming completely to avoid the issues identified above \cite{11}.
The ``no side effects'' advantage of Functional Programming can
be achieved very simply in Object-Orientated by creating all objects
as Immutable by default. This addresses the problems identified above,
as well as providing other advantages \cite{10}:
\begin{itemize}
\item Immutable Objects are thread safe by design.
\item They are simple to create and copy.
\item They do not need to be validated after creating as they do not change.
\item Objects can be made mutable only when needed.
\end{itemize}

\section{Immutable Design Patterns}

\subsection{Proto-Patterns Versus Design Pattern}

The three Immutable Design Patterns defined below are called ``design
patterns'' as part of this paper, but they may be more correctly
called ``proto-patterns'' rather than full Design Patterns. 

The standard definition of a Design Pattern, according to the Portland
Pattern Repository \cite{15}, is that they occur in at ``least three
systems''. The patterns described here may be better described as
``proto-patterns'', defined as ``something documented in a pattern
form, but lacks supporting known uses'' \cite{16}. I leave it to
the reader to decide whether they have seen these patterns in use.
The term Immutable Design Patterns is used with this caveat.

\subsection{Immutable Object Pattern}

\subsubsection{Description}

This design pattern is the simplest of the three as it just describes
the creation of an Immutable Object. Listing 1 illustrates this pattern. 

The class ImmutablePerson has the following properties:
\begin{itemize}
\item The constructor takes all the necessary values to construct the object
completely.
\item There are no mutator methods on the class.
\item All values returned by accessor method return a copy of the instance
variable value, rather than the original value. This is to avoid changes
to attribute values as per Listing 2. 
\item The constructor calls the Ruby freeze method which is somewhat similar
to the final class attribute in Java. Calling freeze on a Ruby object
makes it Immutable. This means that after the constructor has completed
processing the class cannot be changed.
\end{itemize}
\texttt{}
\begin{lstlisting}[caption={Immutable Object Example 1},basicstyle={\footnotesize\ttfamily},language=Ruby,showstringspaces=false]

class ImmutablePerson
  attr_reader :name

  def initialize name, address
	@name = name        
	@address = address     
	 self.freeze
  end

  def address
	return @address.clone   
  end 
end

class Address   
  attr_accessor :line1, :line2, :line3
  def initialize line1, line2, line3     
    @line1 = line1     
    @line2 = line2     
    @line3 = line3 	
  end
end
\end{lstlisting}

\subsubsection{Advantages and Disadvantages}
\begin{itemize}
\item This Design Pattern creates a completely Immutable Object in a simple
way.
\item However, in a situation where you are working with existing Mutable
Objects this Design Pattern will be of no benefit. The next two patterns
will be of more use here.
\end{itemize}
\begin{lstlisting}[caption={Immutable Object Example 2},basicstyle={\footnotesize\ttfamily}]
address =  Address.new("Foo Street", "Bar Town", "Test County") 
person =  ImmutablePerson.new("Foo McBar", address) 
puts person.address.line1             # "Foo Street"
person.address.line1 = "Bar Street" 
puts person.address.line1             # still "Foo Street"

\end{lstlisting}

\subsection{Immutable Subclass Pattern}

\subsubsection{Description}

Listing 3 illustrates this pattern. This Design Pattern illustrates
a situation where a Mutable class has a Mutable and an Immutable subclass.
\begin{itemize}
\item The class AbstractPerson is an abstract superclass. 
\item AbstractPerson has two subclasses - MutablePerson, a standard Mutable
Object. ImmutablePerson is a Immutable Object.
\item ImmutablePerson calls the constructor on the superclass and then freezes
the class.
\item Rather than passing a copy of the attribute (as in Listing 1) to avoid
changes to attribute values, this example freezes the instance variable
object. This achieves the same outcome as illustrated in Listing 4.
This may be more useful in some situations where an instance variable
value is difficult to clone (memory resources etc.). This will also
throw an error when changes to instance variables are attempted. This
error may be useful in debugging unwanted changes to state.
\end{itemize}
\begin{lstlisting}[caption={Immutable Subclass Pattern Example 1},basicstyle={\footnotesize\ttfamily}]
class AbstractPerson
  attr_accessor :name, :address
  def initialize name, address     
    @name = name     
    @address = address   
  end
end

class Address   
  attr_accessor :line1, :line2, :line3
  def initialize line1, line2, line3     
    @line1 = line1     
    @line2 = line2     
    @line3 = line3   
  end
end

class MutablePerson < AbstractPerson   
  # additional behaviour here... 
end

class ImmutablePerson < AbstractPerson
  def initialize name, address     
    super name, address     
    self.freeze     
    @address.freeze  # freeze inst vars
   end

  def address     
    return @address  
  end
end
\end{lstlisting}

\subsubsection{Advantages and Disadvantages}
\begin{itemize}
\item This Design Pattern allows a system to differentiate between Mutable
and Immutable objects using inheritance. For instance this may be
useful where part of the object hierarchy are needed only for reporting
and can be read only.
\item In some situations inheritance may not be the best way to differentiate
between Mutable and Immutable objects. In these situations the next
Design Pattern would be more beneficial.
\end{itemize}
\begin{lstlisting}[caption={Immutable Subclass Pattern Example 2},basicstyle={\footnotesize\ttfamily}]
address =  Address.new("Foo Street", "Bar Town", "Test County")
person1 =  MutablePerson.new("Foo McBar", address) 
person1.address.line1 = "Bar Street" 
puts person1.address.line1       # "Bar Street"
person2 =  ImmutablePerson.new("Bar McFoo", address) 
person2.address.line1 = "Bar Street"  # error: can't modify frozen
\end{lstlisting}

\subsection{Immutable Adapter Pattern}

\subsubsection{Description}

Listing 5 illustrates this pattern. This Design Pattern illustrates
a situation where a Mutable class is wrapped in an Immutable class.
\begin{itemize}
\item The class Person is a standard Mutable class. 
\item ImmutablePerson contains an instance of Person in an instance variable.
\item The constructor of ImmutablePerson constructs a new Person class and
then freezes the objects and it's instance variable values.
\item The accessors of ImmutablePerson query the wrapped Person class. As
per Listing 6, this example freezes the instance variable objects.
This will also throw an error when changes to instance variables are
attempted. 
\end{itemize}
\begin{lstlisting}[caption={Immutable Adapter Pattern Example 1},basicstyle={\footnotesize\ttfamily}]
class Person
  attr_accessor :name, :address
  def initialize name, address     
    @name = name     
    @address = address   
  end
end

class Address   
  attr_accessor :line1, :line2, :line3
  def initialize line1, line2, line3
    @line1 = line1     
    @line2 = line2     
    @line3 = line3   
  end
end

class ImmutablePerson
  def initialize name, address     
   @mutable_person = Person.new name, address     
   @mutable_person.freeze     
   @mutable_person.address.freeze   
  end

  def name     
    return @mutable_person.name   
  end

  def address     
    return @mutable_person.address   
  end
end
\end{lstlisting}

\subsubsection{Advantages and Disadvantages}
\begin{itemize}
\item This Design Pattern allows any Mutable Object to be wrapped and passed
around as an Immutable Object.
\item This Design Pattern, in keeping with the Adapter Design Pattern, will
allow the Mutable Object to be extended if necessary.
\item This is the most useful of the three Immutable Design Pattern in existing
systems but does involve indirection. For simplicity of understanding
the other two Immutable Design Patterns should be considered first.
\end{itemize}
\begin{lstlisting}[caption={Immutable Adapter Pattern Example 2},basicstyle={\footnotesize\ttfamily}]
address =  Address.new("Foo Street", "Bar Town", "Test County")
person1 =  Person.new("Foo McBar", address) 
person1.address.line1 = "Bar Street" 
puts person1.address.line1            # "Bar Street"
person2 =  ImmutablePerson.new("Bar McFoo", address) 
person2.address.line1 = "Bar Street"  # error: can't modify frozen

\end{lstlisting}

\section{Related Work }

There has been no formal definition of Immutable Design Patterns in
Ruby as far as the author is aware but there is plenty information
on Immutability across the web:
\begin{itemize}
\item The Portland Pattern Repository \cite{17} contains much discussion
on Mutable and Immutable Objects across many programming languages.
\item Hamster is a set of ``Efficient, Immutable, Thread-Safe Collection
classes for Ruby''. All Hamster collections are Immutable making
them thread safe \cite{18}.
\item The Functional Ruby blog offers a wider perspective on Functional
Programming in Ruby \cite{19}.
\end{itemize}

\section{Summary }

In this paper, the three Immutable Design Patterns in Table 1 were
discussed. Implementing these Design Patterns will allow a programmer
to take advantage of the ``no side-effects'' properties of Functional
Programming within an Object-Orientated programming language such
as Ruby.

\begin{table}

\begin{tabular}{|c|c|}
\hline 
Name & Description\tabularnewline
\hline 
\hline 
Immutable Object Pattern & Object itself is immutable\tabularnewline
\hline 
Immutable Subclass Pattern & Immutable subclass of mutable parent class\tabularnewline
\hline 
Immutable Adapter Pattern & Immutable wrapper around mutable class\tabularnewline
\hline 
\end{tabular}

\caption{Summary of Ruby Immutable Design Patterns}

\end{table}

\end{document}